\documentclass[prb,onecolumn,notitlepage,superscriptaddress]{revtex4-1}

\usepackage[dvips]{graphicx}
\usepackage{amssymb,amsfonts,amsmath,color}

\newcommand{\xhat}{\hat {\bf x}}
\newcommand{\bs}{{\bf {s}}}
\newcommand{\br}{{\bf {r}}}
\newcommand{\bx}{{\bf {x}}}
\newcommand{\bv}{{\bf {v}}}
\newcommand{\bk}{{\bf {k}}}
\newcommand{\bdr}{{\bf {dr}}}

\newcommand{\bom}{{\mbox{\boldmath $\omega$}}}

\graphicspath{Figures_for_paper/}
\begin{document}

\title{Local and nonlocal dynamics in superfluid turbulence}

\author{L.~K.~Sherwin-Robson}
\author{C.~F.~Barenghi}
\affiliation{Joint Quantum Centre Durham-Newcastle, 
School of Mathematics and Statistics, University of Newcastle, 
Newcastle upon Tyne, NE1 7RU, UK}
\author{A.~W.~Baggaley}
\affiliation{School of Mathematics and Statistics, 
University of Glasgow, Glasgow, G12 8QW, UK}

\begin{abstract}
In turbulent superfluid He II, the
quantized vortex lines interact via the classical Biot--Savart law to
form a complicated vortex tangle.
We show that vortex tangles with the same vortex line density will have 
different energy spectra, depending on the normal fluid which feeds 
energy into the superfluid component, and
identify the spectral signature of two forms of superfluid turbulence:
Kolmogorov tangles and Vinen tangles.
By decomposing the superfluid velocity field into local and nonlocal
contributions, we find that in Vinen tangles the motion of vortex
lines depends mainly on the local curvature, whereas in Kolmogorov
tangles the long-range vortex interaction is dominant and leads to
the formation of clustering of lines, in analogy to the 'worms` of
ordinary turbulence.

\end{abstract}

%\keywords{quantum turbulence \and quantized vortices \and coherent
%structures}

\maketitle

\section{Introduction}
\label{section:1}

The hydrodynamics of helium~II is noteworthy for two reasons: its two-fluid
nature (an inviscid superfluid and a viscous normal fluid),
and the fact that superfluid vorticity is constrained
to thin, discrete vortex lines of fixed (quantized)
circulation \cite{Donnelly}; 
in ordinary (classical) fluids, by contrast, the vorticity 
is a continuous field. 
Turbulence in helium~II (called superfluid turbulence, or quantum
turbulence)  consists of a three-dimensional
tangle of  interacting vortex lines. The properties of this new
form of turbulence and current thinking (in terms of theory and
experiments) have been recently reviewed 
\cite{Barenghi-Skrbek-Sreeni,Nemirovskii}.   

Under certain conditions, it has been argued 
\cite{Vinen-Niemela,Skrbek-Sreeni} that 
the turbulent tangle is characterized by a single length scale,
the average distance $\ell$ between the vortex lines, which is inferred
from the experimentally observed vortex line density $L$ (length of vortex
line per unit volume) as $\ell \approx L^{-1/2}$. 
Models based on this property describe fairly well 
the pioneering experiments of Vinen \cite{Vinen}, 
in which an applied heat flux drives the superfluid and the normal fluid in 
opposite directions (thermal counterflow). 
More recently, such `Vinen' tangles were created at very low temperatures 
by short injections of ions \cite{Walmsley-Golov}, exhibiting the 
characteristic decay $L \sim t^{-1/2}$ predicted by Vinen 
\cite{Baggaley-ultraquantum}. 

Under different conditions, however, the experimental evidence is
consistent with a more structured vortex tangle
\cite{Vinen-Niemela,Volovik},
where the kinetic energy is distributed over
a range of length scales according to the same Kolmogorov law which governs
ordinary turbulence. `Kolmogorov' tangles have been
generated at high temperatures by stirring liquid helium with grids
\cite{Donnelly-grid} or propellers \cite{Tabeling,Salort}, 
and at very low temperatures by an intense injection
of ions \cite{Walmsley-Golov}, exhibiting the decay $L \sim t^{-3/2}$
expected from the energy spectrum 
\cite{Donnelly-grid,Baggaley-ultraquantum}. 

The experimental 
evidence for these two forms of superfluid turbulence is only indirect and
arises from macroscopic observables averaged over the experimental 
cell, such as
pressure \cite{Tabeling,Salort} and
vortex line density \cite{Donnelly-grid},
not from direct visualization of vortex lines.
In a recent paper \cite{Sherwin} we have characterized the energy spectrum
of the two forms of turbulence, and showed
that 'Kolmogorov' turbulence contains metastable, 
coherent vortex structures 
\cite{Baggaley-structures,Baggaley-Laurie}, 
similar perhaps to the `worms' which are
observed in ordinary turbulence \cite{Frisch}. 
The aim of this work is to go a step
further, and look for the dynamical origin of the reported spectral difference
and coherent structures.

It is well known \cite{Saffman} that in an incompressible fluid
the velocity field $\bv$
is determined by the instantaneous distribution of vorticity $\bom$
via the Biot-Savart law:

\begin{equation}
\bv(\bx)=\frac{1}{4 \pi} \int \frac{\bom(\bx') \times (\bx-\bx')  }
{\vert \bx-\bx' \vert^3   } d^3\bx'
\label{eq:BS1}
\end{equation}
\noindent
where the integral extends over the entire flow.
The question which we address
is whether the velocity at the point $\bx$ is mainly determined
by the (local) vorticity near $\bx$ or by (nonlocal) contributions from
further away. Since the quantization of the circulation implies that
the velocity field around a vortex line is 
strictly $1/r$ (where $r$ is the radial distance from the line), 
from the predominance of local effects we would infer
that the vorticity is randomly distributed and nonlocal effects
cancel each other out; conversely, the predominance of nonlocal
effects would suggest the existence of coherence structures. 

If the vorticity were a continuous field, the
distinction between local and nonlocal would involve 
an arbitrary distance, however in our problem the concentrated nature of vorticity
introduces a natural distinction between local and nonlocal
contributions, as we shall see.

\section{Method}
\label{section:2}
The two most popular models \cite{Barenghi-Skrbek-Sreeni}
for studying superfluid turbulence are the 
Gross-Pitaevskii equation (GPE) and the Vortex Filament Model (VFM).
Each has advantages and disadvantages. The main advantage of the GPE
over the VFM is that vortex reconnections are solutions of the equation
of motion and do not require an ad--hoc algorithmic procedure. On the
other hand, the experimental context of our interest is liquid helium at 
intermediate temperatures where the effects of the friction are 
important (at the chosen value $T=1.9~K$, typical of experiments,
the normal fluid fraction is 42 percent). We must keep in mind that,
although the GPE is a good quantitative
model of weakly interacting atomic gases, it is only an idealized model
of liquid helium; moreover, the GPE applies only at very low temperatures,
and there is not yet a consensus\cite{Proukakis2008} 
on its finite-temperature generalizations.
This is why, following the approach of Schwarz\cite{Schwarz},
we choose to use the VFM and model superfluid vortex
lines as space curves $\bs(\xi,t)$ (where $t$ is time and $\xi$ is arc length) 
of infinitesimal thickness and circulation 
$\kappa=9.97 \times 10^{-4}~\rm cm^2/s$ which move according to

\begin{equation}
\label{eq:Schwarz}
\frac{ds}{dt} = \bv_s + \alpha \bs' \times (\bv_n^{ext} - \bv_s) 
- \alpha' \bs' \times [\bs' \times (\bv_n^{ext} - \bv_s)].
\end{equation}

\noindent
Here $\alpha$ and $\alpha'$ are
temperature-dependent friction coefficients \cite{BDV,DB}, $\bv_n$ is
the normal fluid velocity, and a prime denotes derivative with respect
to arc length (hence $\bs'=d\bs/d\xi$ is the local unit tangent vector, and
$C=\vert \bs''\vert$ is the local curvature; the radius of curvature,
$R=1/C$, is the radius of the osculating circle at the point $\bs$).
The superfluid velocity consists of two parts: $\bv_s=\bv_s^{ext}+\bv_s^{self}$. 
The former $\bv_s^{ext}$ represents any externally applied superflow; 
the latter $\bv_s^{self}$ the self-induced velocity 
at the point $\bs$, results from Equation~\ref{eq:BS1} in the limit of 
concentrated vorticity:

\begin{equation}
\bv_s^{self}(\bs)=
-\frac{\kappa}{4\pi} \oint_{\mathcal{L}} 
\frac{(\bs - \br)}{|\bs - \br|^3} \times \bdr,
\label{eq:BS2}
\end{equation}

\noindent
where the line integral extends over the entire vortex 
configuration $\mathcal{L}$.

We discretised the vortex lines into a large number of points 
$\bs_j$ ($j=1, \cdots N$). The minimal separation $\delta$ between
the points is such that
the vortex curves are sufficiently smooth (at the temperatures
of interest here, the friction, controlled by
the parameters $\alpha$ and $\alpha'$, damps out high frequency
perturbations called Kelvin waves).
The VFM assumes that vortex lines are infinitely thin, thus
Eq.~(\ref{eq:BS2}) is valid only if $\vert \bs - \br \vert >> a_0$
where $a_0$ is the vortex core (the region around the vortex
axis where the superfluid density drops from its bulk value to zero).
Thus the integral diverges if one attempts to find the velocity at
$\br=\bs$. As remarked by Schwarz, the occurrence of a similar problem in 
classical hydrodynamics is not helpful, since the physics of the vortex
core is different. The solution to the problem which was proposed by Schwarz
\cite{Schwarz1985} and thereafter adopted in the helium literature
is based on Taylor expanding the integrand around the
singularity and comparing against the well-know expression for the
 self-induced velocity of a circular ring. In this way he obtained
a decomposition of the
self-induced velocity
at the point $\bs_j$ (Eq.~\ref{eq:BS2}) into the following local 
and nonlocal contributions:

\begin{equation}
\bv_s^{self}(\bs_j)=\bv_s^{loc}(\bs_j)+\bv_s^{non}(\bs_j)=
\frac{\kappa}{4\pi} \ln \left( \frac{\sqrt{\Delta \xi_{+}\Delta \xi_{-}}}{a_0} \right) 
\bs'_j \times \bs''_j 
+ \frac{\kappa}{4\pi} \oint_{\mathcal{L}'} \frac{(\bs_j - \br)}{|\bs_j - \br|^3} \times \bdr
\label{eq:BS3}
\end{equation}

\noindent
where $\Delta \xi_{+}$ and $\Delta \xi_{-}$ are the arc lengths of 
the curves between the point $\bs_j$ and the adjacent points $\bs_{j-1}$ 
and $\bs_{j+1}$ along the vortex line.  $\mathcal{L}'$ is the original 
vortex configuration $\mathcal{L}$ but now without the 
section between $\bs_{j-1}$ and $\bs_{j+1}$.
The superfluid vortex core radius $a_0 \approx 10^{-8}~\rm cm$ 
acts as cutoff parameter. Details and tests of the numerical techniques 
against the experimental and the numerical literature
are published elsewhere 
\cite{Baggaley-reconnections,Baggaley-stats,Baggaley-PNAS,Adachi}. 
Note that the local contribution
is proportional to $\bs' \times \bs''$, in the binormal direction.

All calculations are performed in a cubic periodic domain of size $D=0.1$
using an Adams-Bashforth time-stepping method (with typical time step
$\Delta t =5 \times 10^{-5}~\rm s$), a tree-method \cite{Baggaley-tree}
with opening angle $\theta=0.4$, and typical minimal resolution 
$\delta=1.6 \times 10^{-3}~\rm cm$.
For example, an increase in the numerical resolution
from $\delta =0.0016$ to $\delta =0.0008$ produces 
a small increase of 2.5\% in the importance of the nonlocal contribution
in Fig.~(\ref{fig:8}). Moreover, the tests against experiments mentioned above 
\cite{Baggaley-stats,Baggaley-PNAS,Adachi}, 
guarantee that the numerical resolution is sufficient, and put the
distinction between $\bv_s^{loc}$ and $\bv_s^{non}$
on solid ground.

It is known from experiments \cite{Bewley2008}
and from more microscopic models \cite{Koplik,Zuccher,Allen2014}
that colliding vortex lines reconnect with each other.
An algorithmic procedure is introduced to
reconnect two vortex lines if they become
sufficiently close to each other. 
This procedure (although arbitrary, unlike the GPE as mentioned
before) has been extensively tested \cite{Baggaley-reconnections}; various
slightly different reconnection algorithms have been proposed and have 
never been found in disagreement with experimental observations.

We choose a temperature typical of experiments, $T=1.9~\rm K$
(corresponding to $\alpha=0.206$ and $\alpha'=0.00834$).
In all three cases,
the initial condition consists of a few seeding vortex loops, 
which interact and reconnect, quickly generating a turbulent vortex 
tangle which appears independent of the initial condition.

\begin{figure}[h]
\begin{center}
\includegraphics[width=0.65\textwidth]{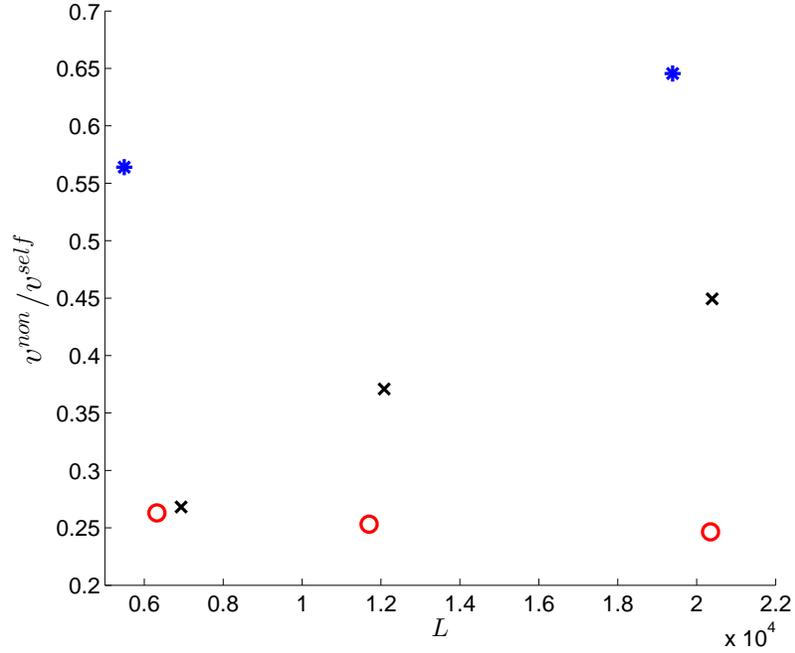}
\caption{(color online).
Ratio $v^{non}/v^{self}$ as a function of vortex line density $L$
$(\rm cm^{-2})$ corresponding to model 1 (uniform normal flow, red circles),
model 2 (synthetic normal flow turbulence, black crosses) and model 3
(frozen Navier-Stokes equation, blue stars).
}

\label{fig:11}
\end{center}
\end{figure}

We study the following
three different regimes of superfluid turbulence, 
characterized by the following forms of the normal 
fluid's velocity field $\bv_n^{ext}$:

\begin{enumerate}
\item{\bf Uniform normal flow.}
Firstly, to model turbulence generated by a small heat flux at the blocked
end of a channel (thermal counterflow), we
impose a uniform normal fluid velocity $\bv_n^{ext}=V_n \xhat$ in the
x-direction (which we interpret as the direction of the channel) 
which is proportional to the applied heat flux; 
to conserve mass,
we add a uniform superflow $\bv_s^{ext}= -(\rho_n/\rho_s) V_n \xhat$ 
in the opposite
direction, where $\rho_n$ and $\rho_s$ are respectively 
the normal fluid and superfluid
densities.  Eqs.~(\ref{eq:Schwarz}) and (\ref{eq:BS3}) are solved in the 
imposed superflow's reference frame. This model is the most used in 
the literature, from
the pioneering work of Schwarz \cite{Schwarz} to the recent
calculations of Tsubota and collaborators \cite{Adachi}.

\item{\bf Synthetic turbulence.}
To model turbulence generated by pushing
helium through pipes or channels \cite{Salort} using plungers or bellows
or by stirring it with grids \cite{Donnelly-grid} 
or propellers \cite{Tabeling}, we start from the
observation that, due to liquid helium's small viscosity $\mu$,
the normal fluid's Reynolds number ${\rm Re}=V D/\nu_n$ is usually large 
(where $V$ is the rms velocity
and $\nu_n=\mu/\rho_n$ the kinematic viscosity), hence we expect the normal
fluid to be turbulent. We assume
$\bv_s^{ext}={\bf 0}$ and \cite{Osborne-2006}

\begin{equation}
\bv_n^{ext}(\bs,t)=\sum_{m=1}^{M}({\bf A}_m \times {\bf k}_m \cos{\phi_m}
+{\bf B}_m \times {\bf k}_m \sin{\phi_m}),
\label{eq:KS}
\end{equation}

\noindent
where $\phi_m=\bk_m \cdot \bs + f_m t$, $\bk_m$ are wavevectors and
$f_m=\sqrt{k^3_m E(k_m)}$ are angular frequencies. 
The random parameters ${\bf A}_m$,  ${\bf B}_m$  and ${\bf k}_m$ are chosen
so that the normal fluid's energy spectrum obeys Kolmogorov's scaling
$E(k_m)\propto k_m^{-5/3}$ in the inertial range $k_1<k< k_M$,
where $k_1 \approx 2 \pi/D$ and $k_M$ correspond to the outer scale of
the turbulence and the dissipation 
length scale respectively. Then we define the Reynolds
number via ${\rm Re}=(k_M/k_1)^{4/3}$.
The synthetic turbulent flow defined by Eq.~(\ref{eq:KS})
is solenoidal, time-dependent,
and compares well with Lagrangian statistics obtained in experiments and
direct numerical simulations of the Navier-Stokes equation. It is therefore
physically realistic and numerically convenient to model current experiments
on grid or propeller generated superfluid turbulence.

\item{\bf Frozen Navier-Stokes turbulence.}
Synthetic turbulence, being essentially the superposition of
random waves, lacks the intense regions of concentrated vorticity which
are typical of classical turbulence \cite{Frisch}. For this reason we consider
a third model: a turbulent flow $\bv_n^{ext}$ obtained by direct numerical
simulation (DNS) of the classical Navier-Stokes equation in a periodic
box with no mean flow. 
Since the simultaneous calculation of superfluid
vortices and turbulent normal fluid would be prohibitively expensive,
we limit ourselves to a time snapshot of $\bv_n^{ext}$. In other words,
we determine the vortex lines under a 'frozen` 
turbulent normal fluid.  Our source for the DNS
data is the John Hopkins Turbulence Database
\cite{JHTDB1,JHTDB2}, which consists of a velocity field on a $1024^3$
spatial mesh. The estimated Reynolds number is 
${\rm Re} \approx (L_0/\eta_0)^{4/3}=3205$ where $L_0$ is the integral scale
and $\eta_0$ the Kolmogorov scale. 
Although ${\rm Re}=3205$ is not a very large Reynolds
number,  at $T=1.9~K$ helium's viscosity is 
$\mu=1.347 \times 10^{-5}~\rm g/(cm~ s)$, the normal fluid density is
$\rho_n=0.0611~\rm g/cm^3$, the kinematic viscosity is
$\nu_n=\mu/\rho_n=0.22 \times 10^{-3}~\rm cm/s$ \cite{DB}, 
and therefore ${\rm Re}=UD/\nu_n=3205$ corresponds to the reasonable 
speed $U=0.7~\rm cm/s$ in a typical $D=1~\rm cm$
channel.
To keep the resulting vortex line density of this model to a computationally
practical value, we rescale the velocity components such that they are
60 percent of their original values, thus obtaining the vortex line density 
$L \approx 20,0000~\rm cm^{-2}$. To obtain Fig.~\ref{fig:11} the scaling factor is only 45 percent, yielding $L \approx 6,000~\rm cm^{-2}$; because of the
nonlinearity of the Navier-Stokes equation, this procedure is clearly an
approximation but is sufficient for our aim of driving a less intense of more intense vortex tangle.

\end{enumerate}

\begin{figure}[h]
\begin{center}
\includegraphics[width=0.45\textwidth]{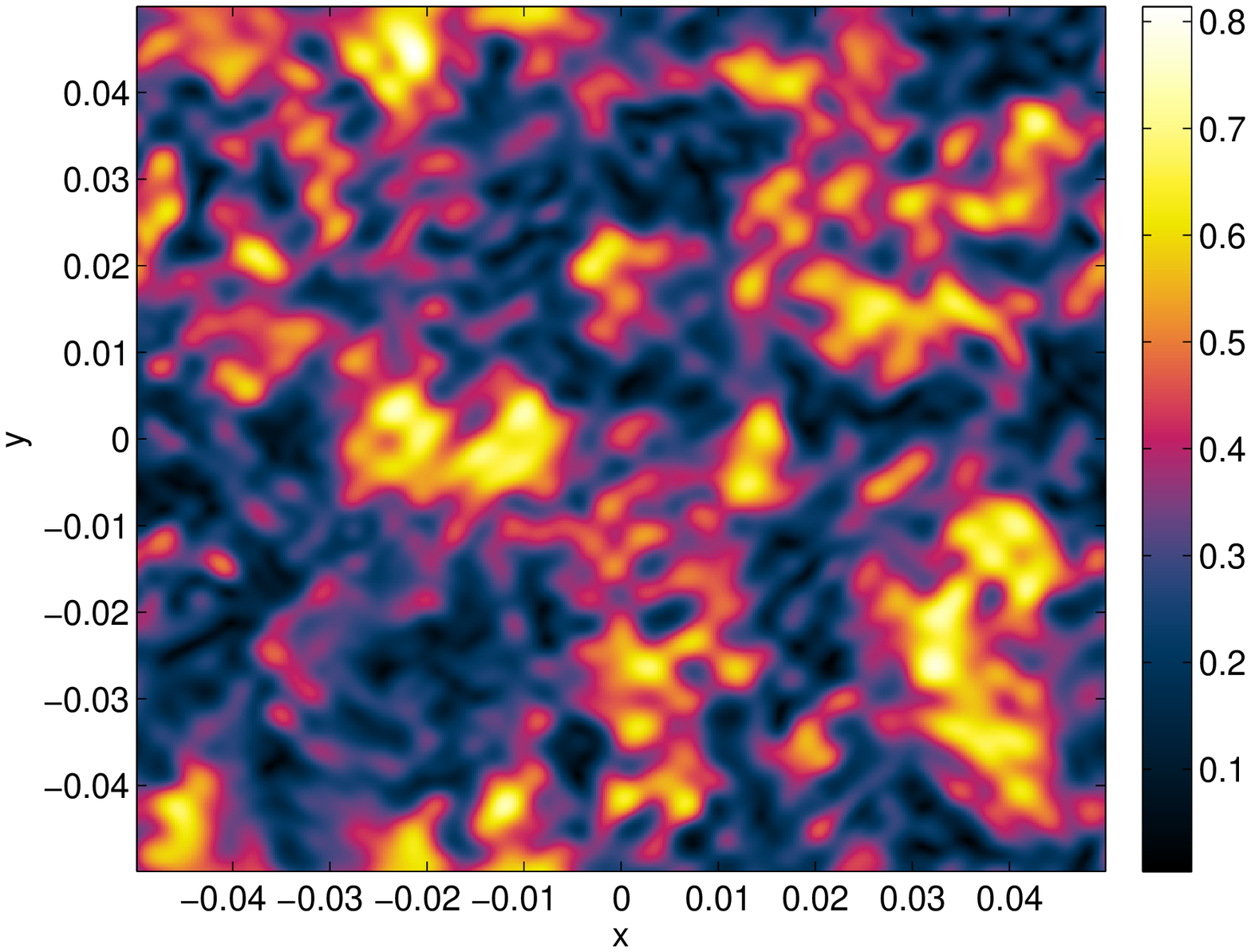}
\includegraphics[width=0.45\textwidth]{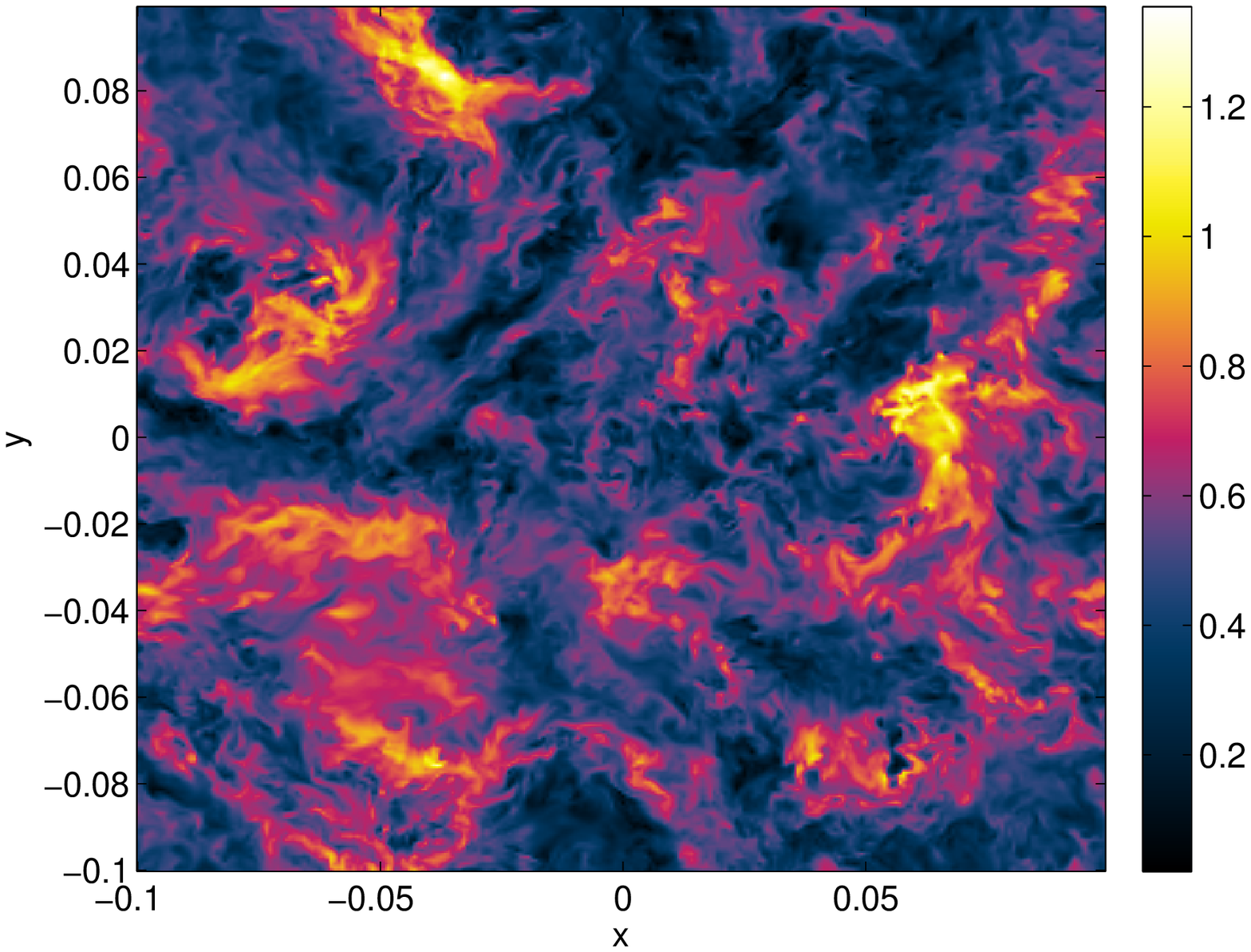}
\caption{(color online).
Magnitude of the driving normal fluid velocity field, 
$\vert \bv_n^{ext} \vert$,
plotted on the $xy$-plane at $z=0$ corresponding to model 2
(synthetic normal flow turbulence, left) and model 3
(frozen Navier-Stokes turbulence, right). The velocity scales
(cm/s) are shown at right of each panel. Note the more localized,
more intense regions of velocity which are present in model 3.
}
\label{fig:1}
\end{center}
\end{figure}

Fig.~\ref{fig:1} shows the magnitude of the normal velocity field plotted
(at a fixed time $t$) on the $xy$ plane at $z=0$,
corresponding to models 2 and 3 (we do not plot the normal fluid velocity
for model 1 because it is uniform). It is apparent that models 1, 2 and 3
represent a progression of increasing complexity of the driving normal flow.
In Fig.~\ref{fig:1}, note in particular the localized regions of 
strong velocity which appear
in model 3.

\section{Results}

The intensity of the turbulence is measured by the vortex line density 
$L=\Lambda/V$ (where $\Lambda$ is the superfluid vortex length in
the volume $V=D^3$), which we 
monitor for a sufficiently long time, such that the properties which we 
report, refer to a statistically steady state of turbulence fluctuating
about a certain average  $L$ (we choose parameters so that $L$ is typical of
experiments).

\begin{figure}[h]
\begin{center}
\includegraphics[width=0.55\textwidth]{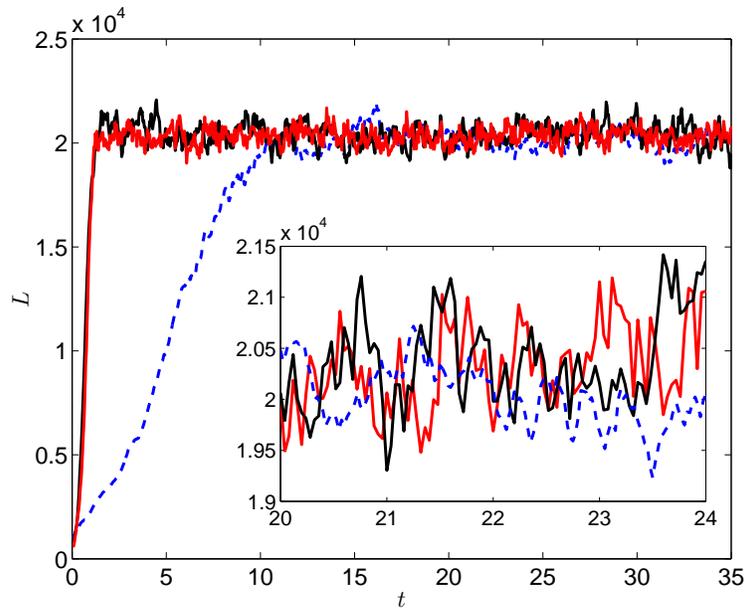}\\
\caption{
(color online).
Evolution of the vortex line density $L$ (cm$^{-2}$) vs time $t$ (s) for
model 1 (red line, uniform normal flow), model 2
(black line, synthetic normal fluid turbulence) and model 3 (dashed blue line,
frozen Navier-Stokes turbulence).
The inset displays the oscillations
of $L$ vs $t$ in more detail. Parameters: temperature $T=1.9~\rm K$,
$V_n=1~\rm cm/s$ (for model 1), ${\rm Re}= 79.44$ (for model 2), and
${\rm Re}= 3025$ (for model 3).
}
\label{fig:2}
\end{center}
\end{figure}

\begin{figure}[h]
\begin{center}
\includegraphics[width=1\textwidth]{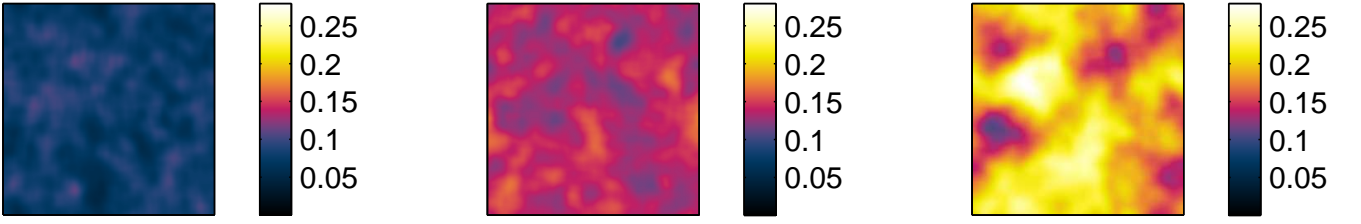}
\caption{
Superfluid energy density
$\epsilon_s=\vert \bv_s  \vert^2/2$ smoothed over the 
average inter-vorex spacing $\ell$, plotted on the $xy$ plane and 
averaged over $z$. Left:  model 1 (uniform normal fluid);
middle: model 2 (synthetic turbulence);
right: model 3 (frozen Navier-Stokes turbulence).
Parameters as in Fig.~(\ref{fig:2}).
}
\label{fig:8}
\end{center}
\end{figure}

Fig.~\ref{fig:2} 
shows the initial 
transient of the vortex line density
followed by the saturation to statistically steady-states
of turbulence corresponding to model 1 (uniform normal flow),
model 2 (synthetic normal flow turbulence) 
and model 3 (frozen Navier-Stokes turbulence).
In all cases, the intensity of the drive is chosen to generate approximately
the same vortex line density, $L \approx 20,000~\rm cm^{-2}$. 

\begin{figure}[h]
\begin{center}
\includegraphics[width=0.30\textwidth]{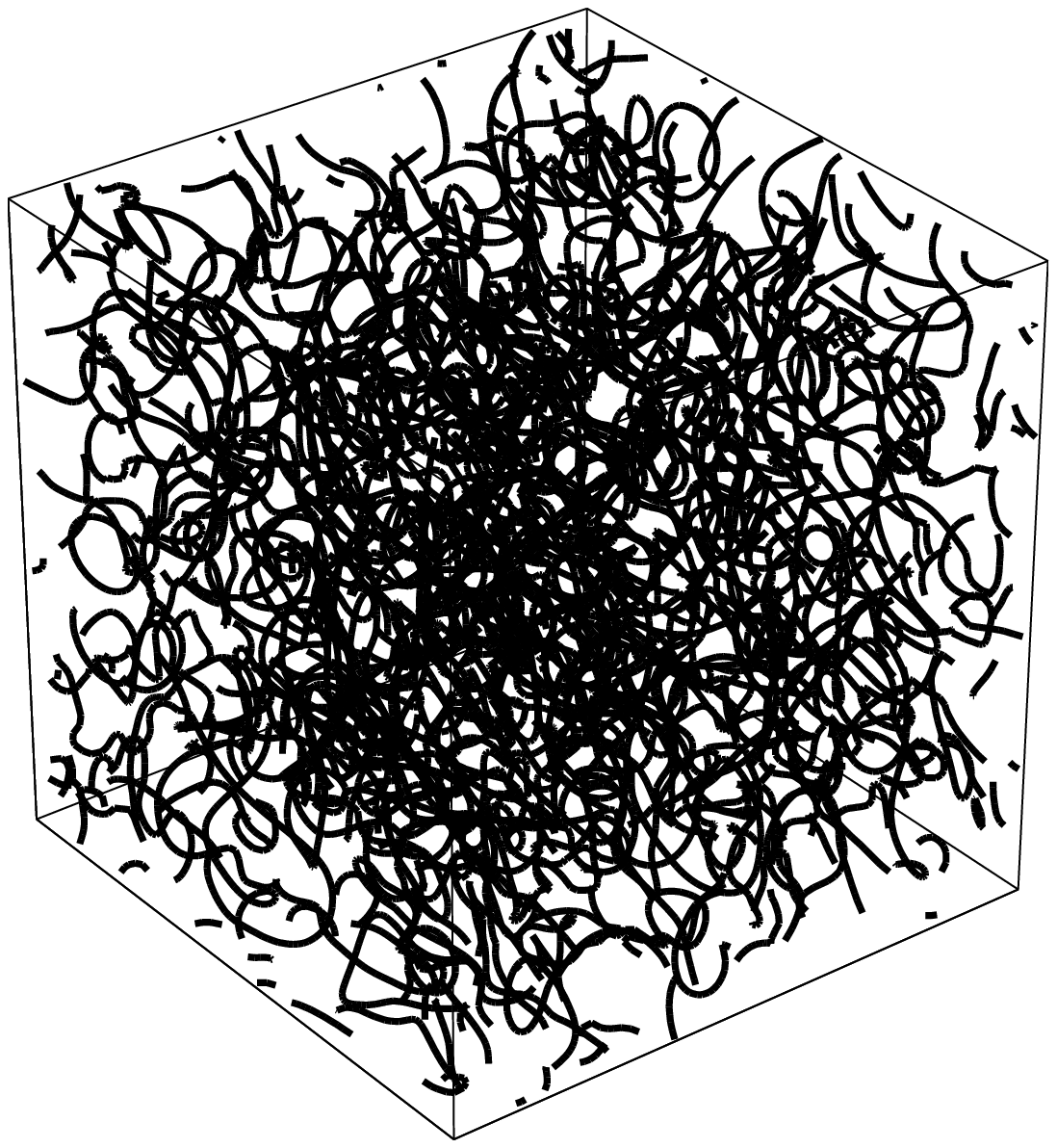}
\includegraphics[width=0.30\textwidth]{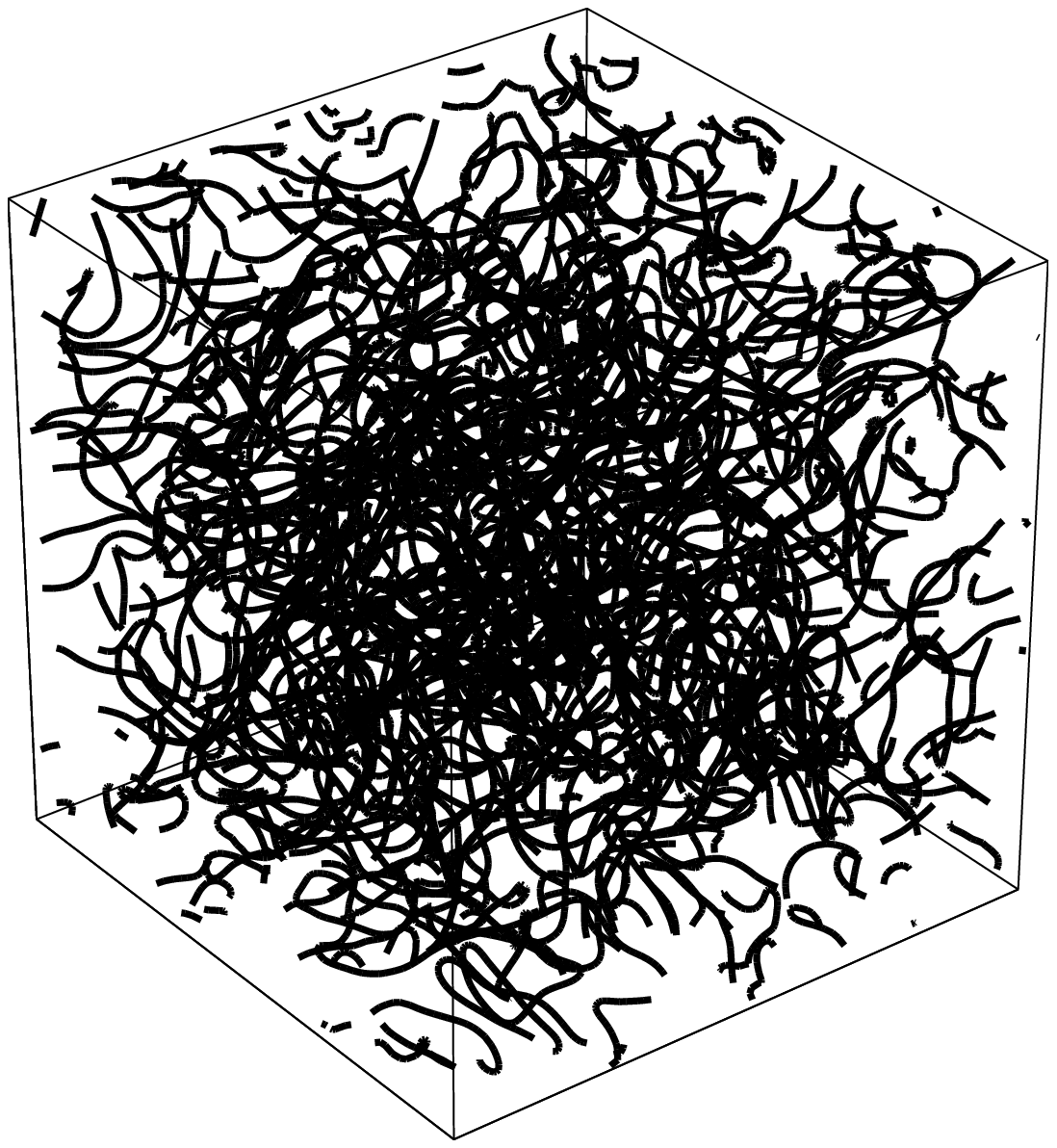}
\includegraphics[width=0.30\textwidth]{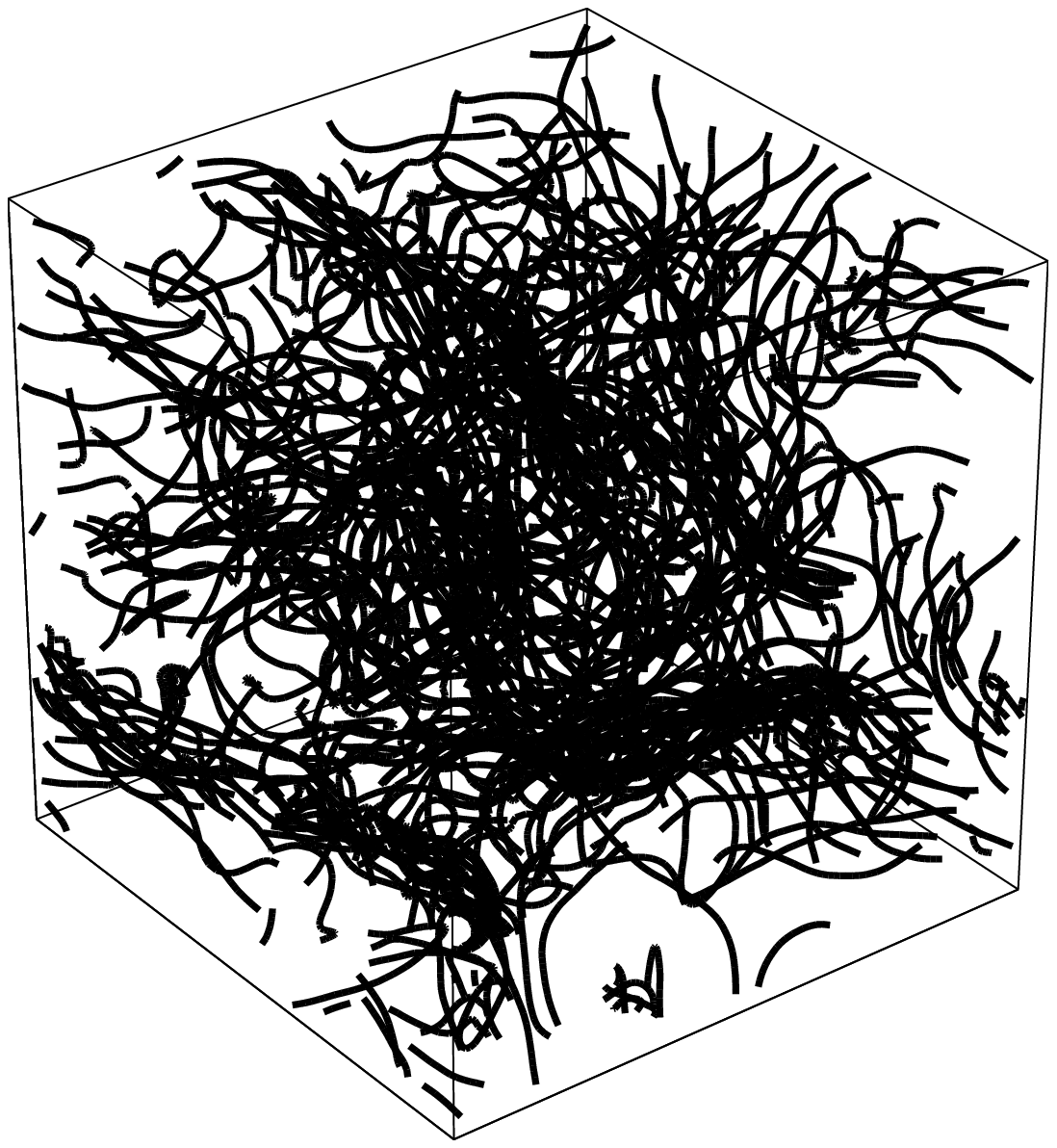}
\caption{
Snapshot of the vortex tangle for model 1 (uniform normal fluid, left),
model 2 (synthetic turbulence, middle) and model 3 (frozen Navier-Stokes
turbulence, right)
at time $t=20~\rm s$ (parameters as in Fig.~(\ref{fig:2}).
}
\label{fig:3}
\end{center}
\end{figure}

Snapshots of the vortex tangles are shown in Fig.~\ref{fig:3}. 
The vortex tangle driven by the uniform normal fluid (model 1, left)
appears visually as the most homogeneous; the vortex tangle driven by the
frozen Navier-Stokes turbulence (model 3, right) appears as the least homogeneous.
What is not apparent in the figure is the mild anisotropy of the
tangle driven by the uniform normal fluid. To quantify this anisotropy,
we calculate the projected vortex lengths in the three Cartesian
directions ($\Lambda_x$, $\Lambda_y$, $\Lambda_z$) and find
$\Lambda_x/\Lambda=0.34 <  \Lambda_y/\Lambda=\Lambda_z/\Lambda=0.54$ for
model 1, confirming a small flattening
of the vortices in the yz plane (this effect was discovered by the early 
investigations of Schwarz \cite{Schwarz}). 
In comparison, models 2 and 3 are more isotropic: for model 2 (tangle driven
by synthetic turbulence) we find
$\Lambda_x/\Lambda=0.50$, 
$\Lambda_y/\Lambda=0.47$ and $\Lambda_z/\Lambda=0.49$),
and for model 3 (tangle driven by frozen 
Navier-Stokes turbulence) we obtain
$\Lambda_x/\Lambda=0.48$, $\Lambda_y/\Lambda=0.50$, $\Lambda_z/\Lambda=0.49$.

\begin{figure}[h]
\begin{center}
\includegraphics[width=0.55\textwidth]{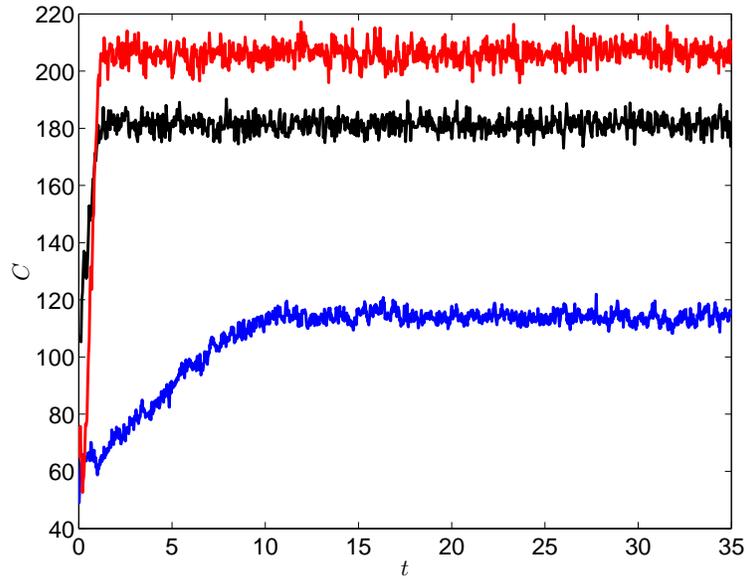}\\
\caption{
(color online).
Average curvature $C$ ($\rm cm^{-1}$) vs time $t$ ($\rm s$)
for model 1 (uniform normal flow, red line), model 2 (synthetic turbulence,
black line) and model 3 (frozen Navier-Stokes turbulence, dashed blue line).
Parameters as in Fig.~(\ref{fig:2}).
}
\label{fig:4}
\end{center}
\end{figure}

\begin{figure}[h]
\begin{center}
\includegraphics[width=0.45\textwidth]{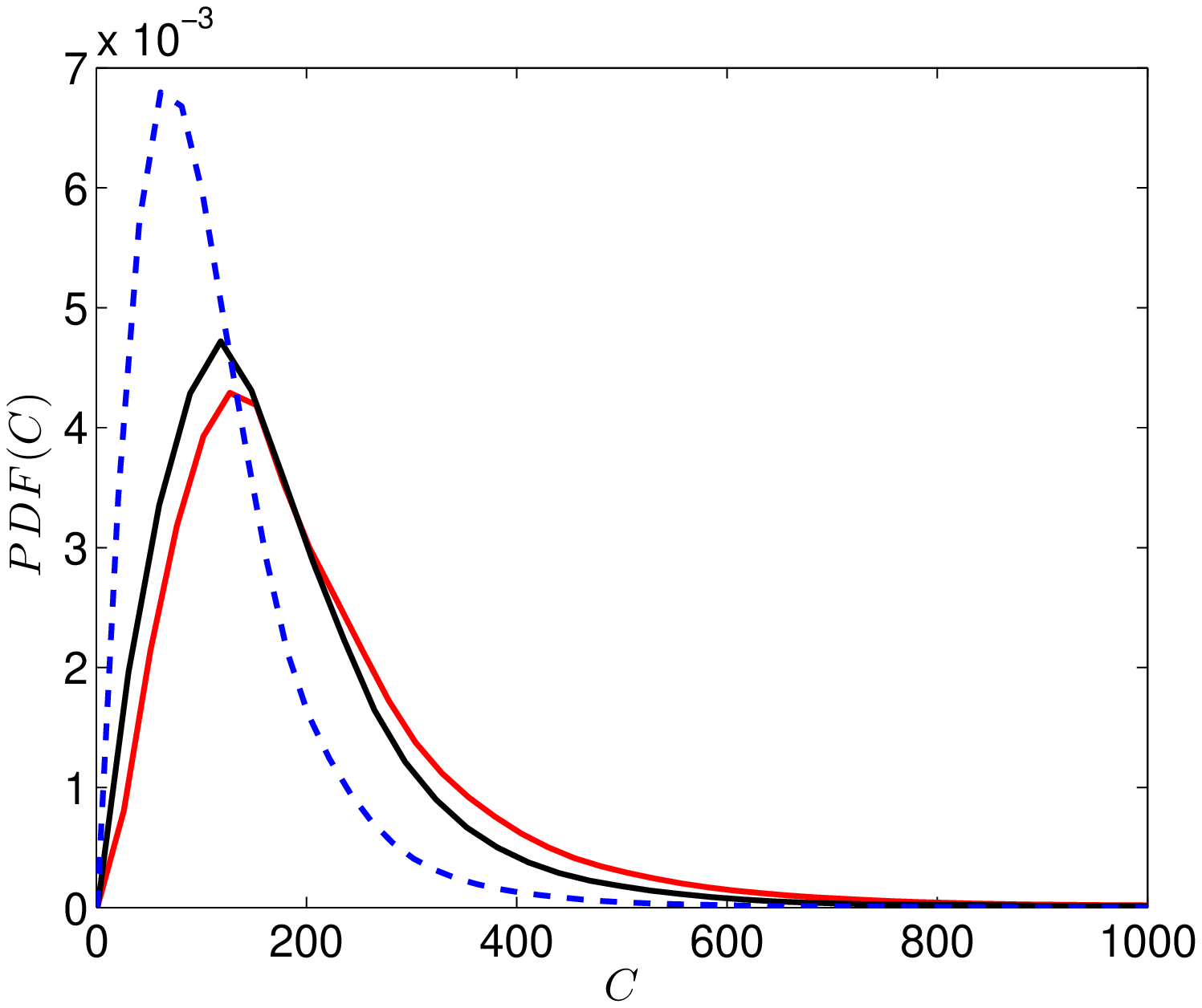}
\includegraphics[width=0.43\textwidth]{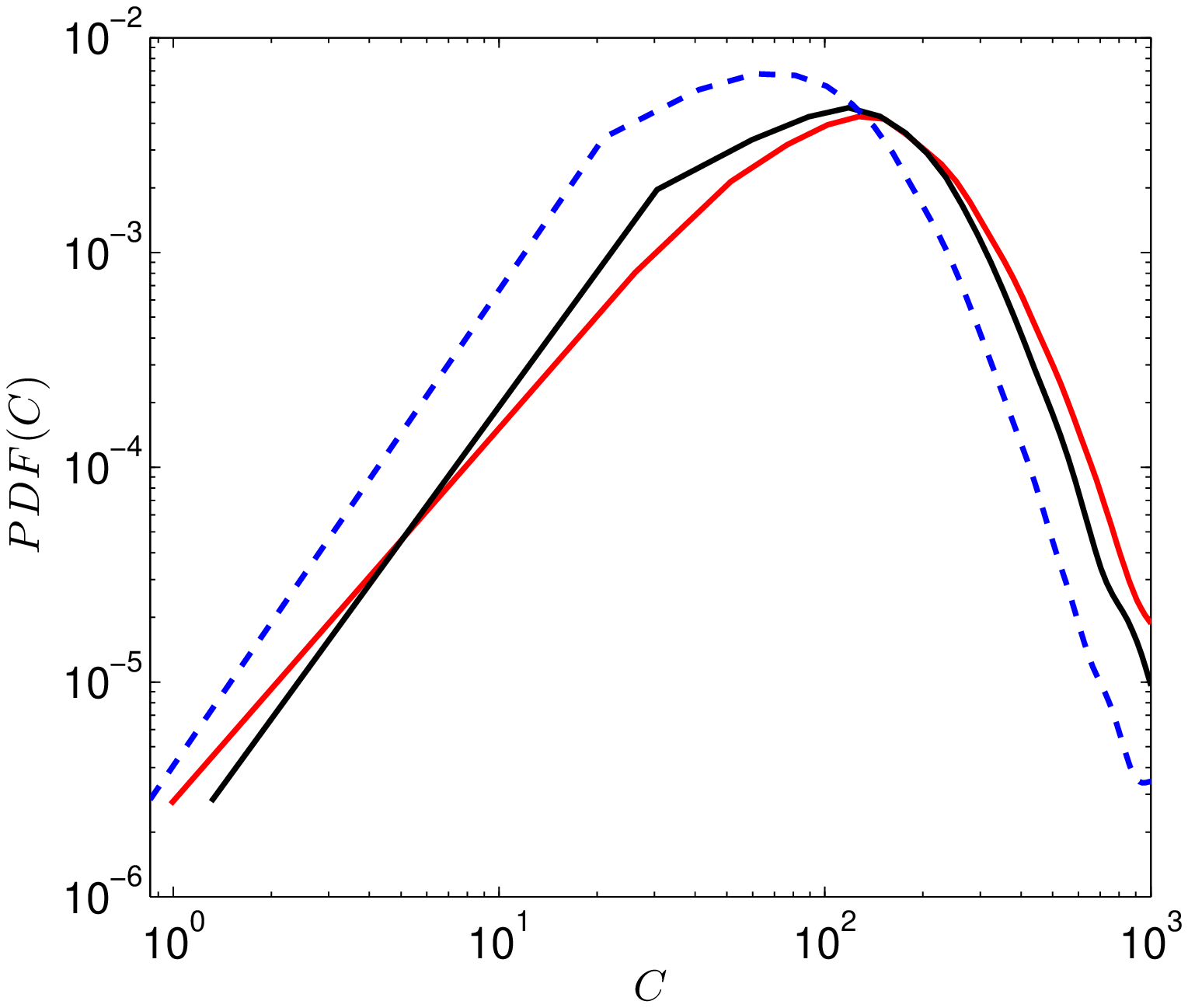}
\caption{
(color online).
Left: Probability density function of the curvature,
${\rm PDF}(C)$, vs curvature, $C$ ($\rm cm^{-1}$), 
corresponding to model 1 (uniform normal flow, red line), model 2
(synthetic turbulence, black line) and model 3
(frozen Navier-Stokes turbulence, dashed blue line). 
Right:  the same data plotted on a log log scale,
where the matching slopes on the plot illustrate that
 we have the same Kelvin waves in all three
models (because they are all at the same temperature).
Parameters as in Fig.~\ref{fig:2}.}
\label{fig:5}
\end{center}
\end{figure}

Fig.~\ref{fig:4} and \ref{fig:5} shows the average curvature
$C=< C_j>$ (sampled over the discretization points $j=1, \cdots N$)
and the distributions of local curvatures $C_j=\vert \bs_j'' \vert$.
The tangle generated by model 3 (frozen Navier-Stokes turbulence) has
the smallest average curvature: the presence of long lines (large radius
of curvature $R=1/\vert \bs'' \vert$) is indeed visible in Fig.~\ref{fig:2}.
In terms of curvature, the tangles generated by models 1 and 2 are more
similar to each other - the average curvature is almost twice as large
as for model 3, indicating that the vortex lines are more in the form
of small loops.

However, vortex line density (Fig.~\ref{fig:2}), visual inspection
(Fig.~\ref{fig:3}) and curvature (Fig.~\ref{fig:4} and \ref{fig:5}) 
do not carry information about the {\it orientation} of
the vortex lines, a crucial ingredient of the dynamics.
Fig.~\ref{fig:6} shows the energy spectrum $E_s(k)$, defined
by
\begin{equation}
\frac{1}{V} \int_V \frac{1}{2}{\bf v}_s^2 dV=\int_0^{\infty} E_s(k) dk
\label{eq:spectrum}
\end{equation}

\noindent
where $k$ is the magnitude of the three-dimensional wavevector.
The energy spectrum describes the distribution of 
kinetic energy over the length scales.
The spectra of the tangles generated by synthetic normal flow
turbulence (model 2)
and by the frozen Navier-Stokes turbulence (model 3) are
consistent with the classical Kolmogorov scaling 
$E_s(k) \sim k^{-5/3}$ for $k < k_{\ell}=2 \pi/\ell$; the kinetic energy
is clearly concentrated at the largest length scales (small $k$).
In contrast,
the spectrum of the tangle generated by the uniform normal fluid (model 1)
peaks at the intermediate length scales, and at large wavenumbers
is consistent with the shallower $k^{-1}$ dependence of individual vortex lines.

A natural question to ask is whether our results are affected
by the particular vortex reconnection algorithm used. In principle, both 
the large $k$ region and the small $k$ region of the spectrum could be affected: 
the former,
because vortex reconnections involve changes of the geometry of the vortices
at small length scales, the latter because the energy flux may be affected.
To rule out this possibility we have performed simulations using the
reconnection algorithm of Kondaurova et al. \cite{Kondaurova},
which tests whether vortex filaments would cross each others path during the
next time step (for details, see also ref.~\cite{Baggaley-reconnections}). 
Fig.~(\ref{fig:7}) is very similar to Fig.~(\ref{fig:6}), confirming
that the shape of the energy spectra does not depend on the 
reconnection algorithm.

It is also instructive to examine the spatial distribution of the superfluid energy
densities arising from the three normal fluid models: Fig.~\ref{fig:8}
displays the superfluid energy density
$\epsilon_s=\vert \bv_s  \vert^2/2$ on the $xy$ plane averaged over $z$.
The left panel (model 1, uniform normal flow) shows that $\epsilon_s$ is
approximately constant, that is to say the vortex tangle is homogeneous;
the middle and right panels (model 2 and 3 for synthetic normal flow
turbulence and frozen Navier-Stokes turbulence) show that the energy
density is increasingly nonhomogeneous, particularly model 3. The localized 
regions of large energy density correspond to vortex lines which are locally 
parallel to each other, reinforcing each other's velocity field rather
than cancelling it out. 

\begin{figure}[h]
\begin{center}
\includegraphics[width=0.65\textwidth]{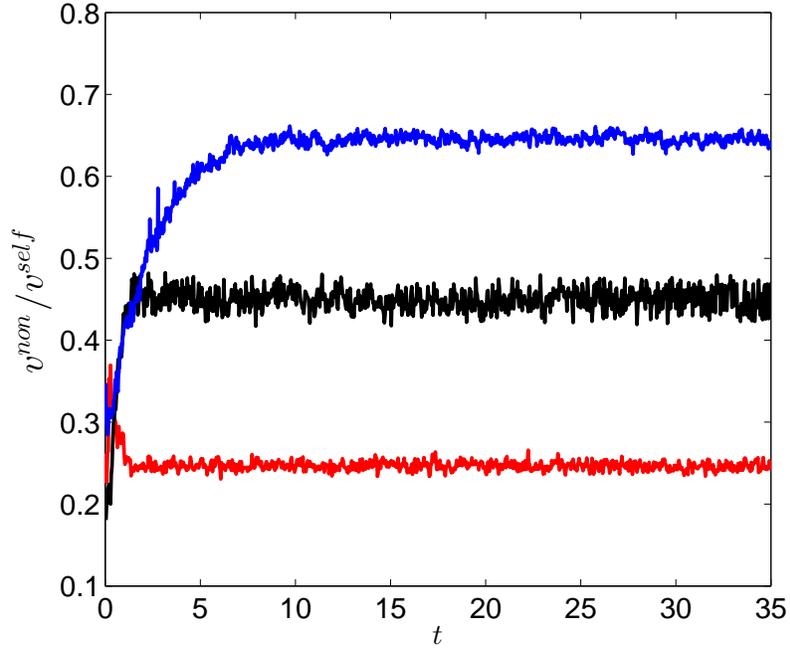}\\
\caption{(color online). Ratio of nonlocal to total self-induced
velocity as a function of time $t$ $\rm (s) $ for tangles generated
by uniform normal fluid (model 1, red line, bottom), synthetic normal fluid
turbulence (model 2, black line, middle) and frozen Navier-Stokes turbulence
(model 3, blue line, top). 
Parameters as in Fig.~(\ref{fig:2}).
}
\label{fig:9}
\end{center}
\end{figure}

The natural question which we ask is what is the cause of the spectral 
difference shown in Fig.~\ref{fig:6}.
To answer the question we examine
the local and nonlocal contributions to the superfluid velocity,
defined according to Equation~\ref{eq:BS3}.
Fig.~\ref{fig:9} shows the fraction $v^{non}/v^{self}$ of the superfluid
velocity which arises from nonlocal contributions, where
$v^{non}=<\vert \bv_{non}(\bs_j)\vert>$ and 
$v^{self}=<\vert \bv_{self}(\bs_j)\vert>$ are sampled over the
discretization points $j=1, \cdots N$ at a given time $t$. 
The difference is striking.
Nonlocal effects are responsible for only 25 percent of the total
superfluid velocity field in the tangle generated by the uniform
normal fluid (model 1), for 45 percent in the tangle generated by
synthetic normal flow turbulence (model 2), 
and for more than 60 percent in the tangle
generated by frozen Navier-Stokes turbulence (model 3).

\begin{figure}[h]
\begin{center}
\includegraphics[width=0.6\textwidth]{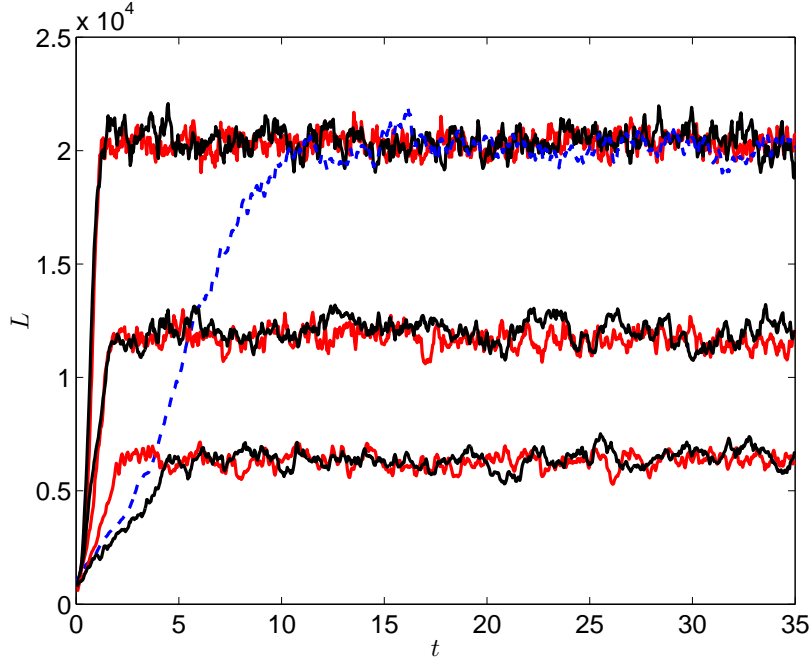}
\caption{
(color online).
Evolution of the vortex line density $L$ (cm$^{-2}$) vs time $t$ (s) 
for model 1 (red line, uniform normal flow, respectively at 
$V_n=1~\rm (cm/s)$ (top),
$V_n=0.75~\rm (cm/s)$ (middle) and $V_n=0.55~\rm (cm/s)$ (bottom)),
model 2 (black line, synthetic normal fluid turbulence, respectively at
$Re=79.44$ (top), $Re=81.59$ (middle) and $Re=83.86$ (bottom)),
and model 3 (dashed blue line, frozen Navier-Stokes turbulence, at $Re=3025$).
}
\label{fig:10}
\end{center}
\end{figure}

Finally, we explore the dependence of the result on the vortex line
density $L$ by generating statistically steady states of turbulence driven
by uniform normal fluid (model 1) and synthetic turbulence (model 2)
with different values of $L$, see Fig.~\ref{fig:10}.
Fig.~\ref{fig:11} shows that for model 1 (uniform normal flow) the
relative importance of nonlocal contributions remains constant
at about 25 percent
over a wide range of vortex line density, from $L \approx 6000$
to $L \approx 20,000~ {\rm cm^{-2}}$, whereas for model 2
(synthetic normal flow turbulence) and 3 (frozen Navier--Stokes turbulence),
it increases with $L$.

\section{Conclusion}

\begin{figure}[h]
\begin{center}
\includegraphics[width=0.420\textwidth]{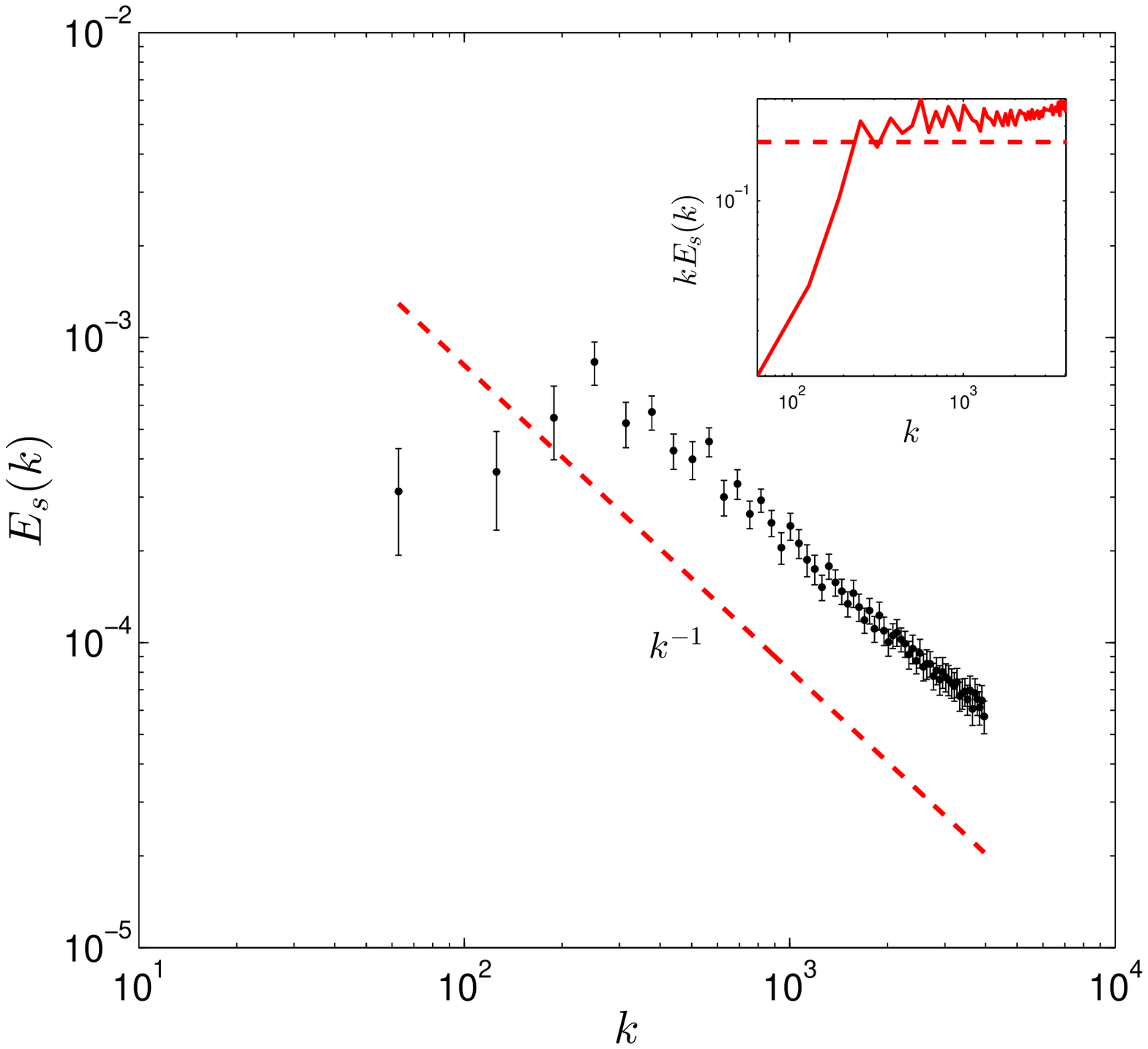}\\
\includegraphics[width=0.420\textwidth]{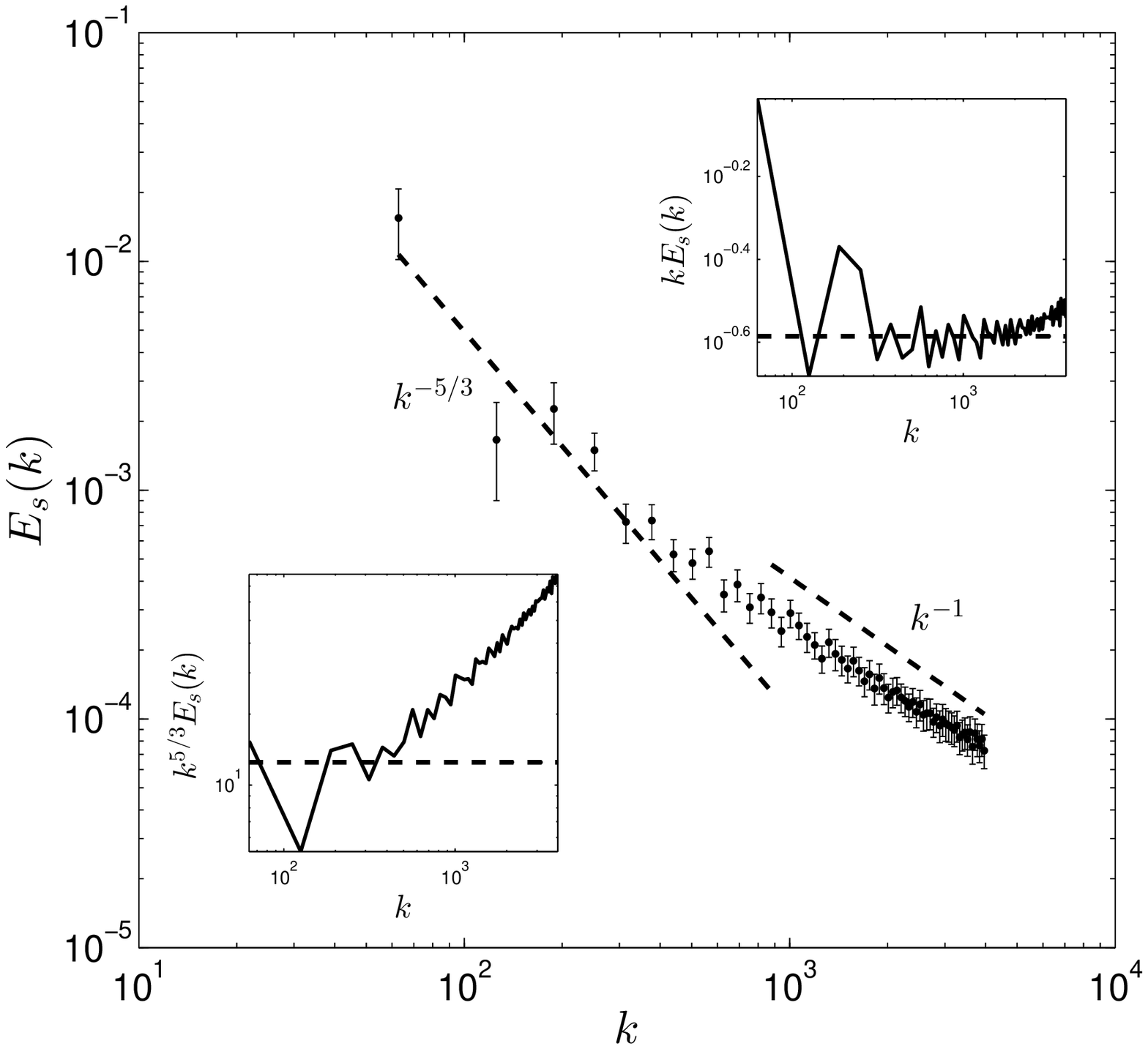} \\
\includegraphics[width=0.420\textwidth]{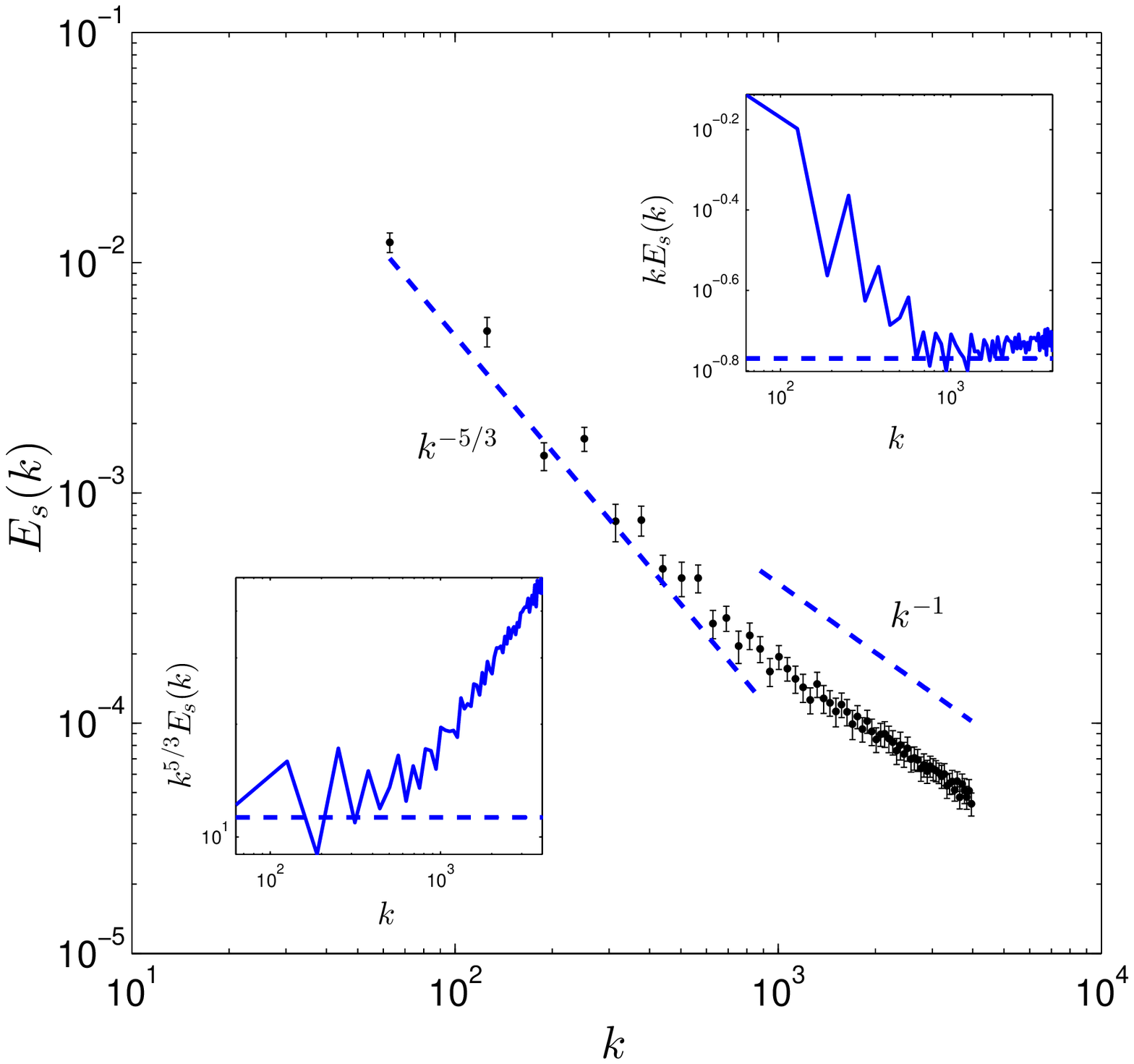}
\caption{
(color online).
Energy spectrum $E(k)$ (arbitrary units) vs wavenumber $k$
($\rm cm^{-1}$) (time averaged over the saturated regime) corresponding
to vortex tangles generated by uniform normal fluid (model 1, top),
synthetic normal fluid turbulence (model 2, middle) and
frozen Navier-Stokes turbulence (model 3, bottom).
The dashed lines indicate the $k^{-1}$ (top) and the $k^{-5/3}$
dependence (middle and bottom), respectively. 
Parameters as in Fig.~(\ref{fig:2}). 
The compensated spectra $k E_s(k)$
and $k^{5/3} E_s(k)$ in the insets show the regions of $k$--space
where the approximate scalings $k^{-1}$ and $k^{-5/3}$ apply.
}
\label{fig:6}
\end{center}
\end{figure}

\begin{figure}[h]
\begin{center}
\includegraphics[width=0.42\textwidth]{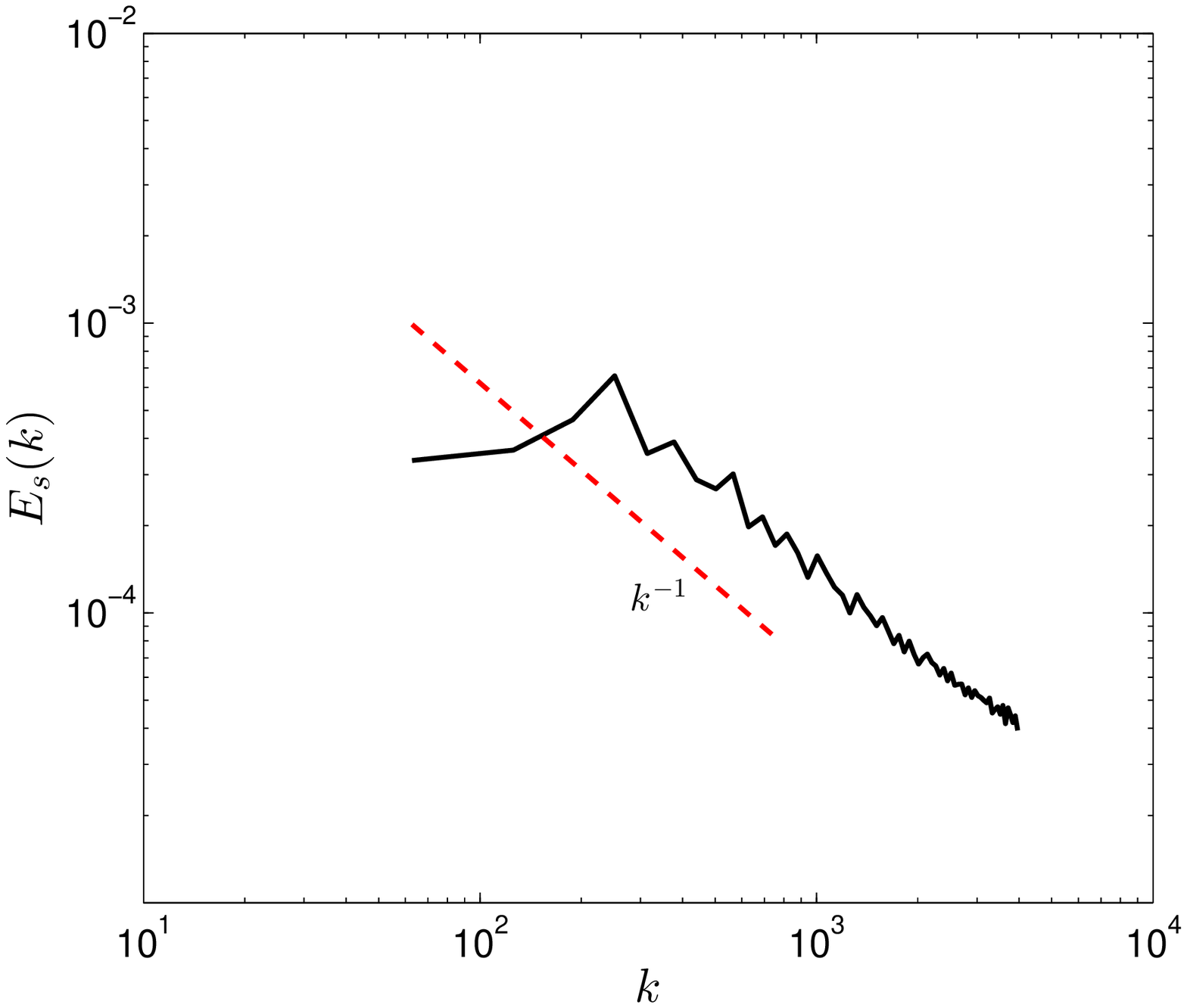}\\
\includegraphics[width=0.42\textwidth]{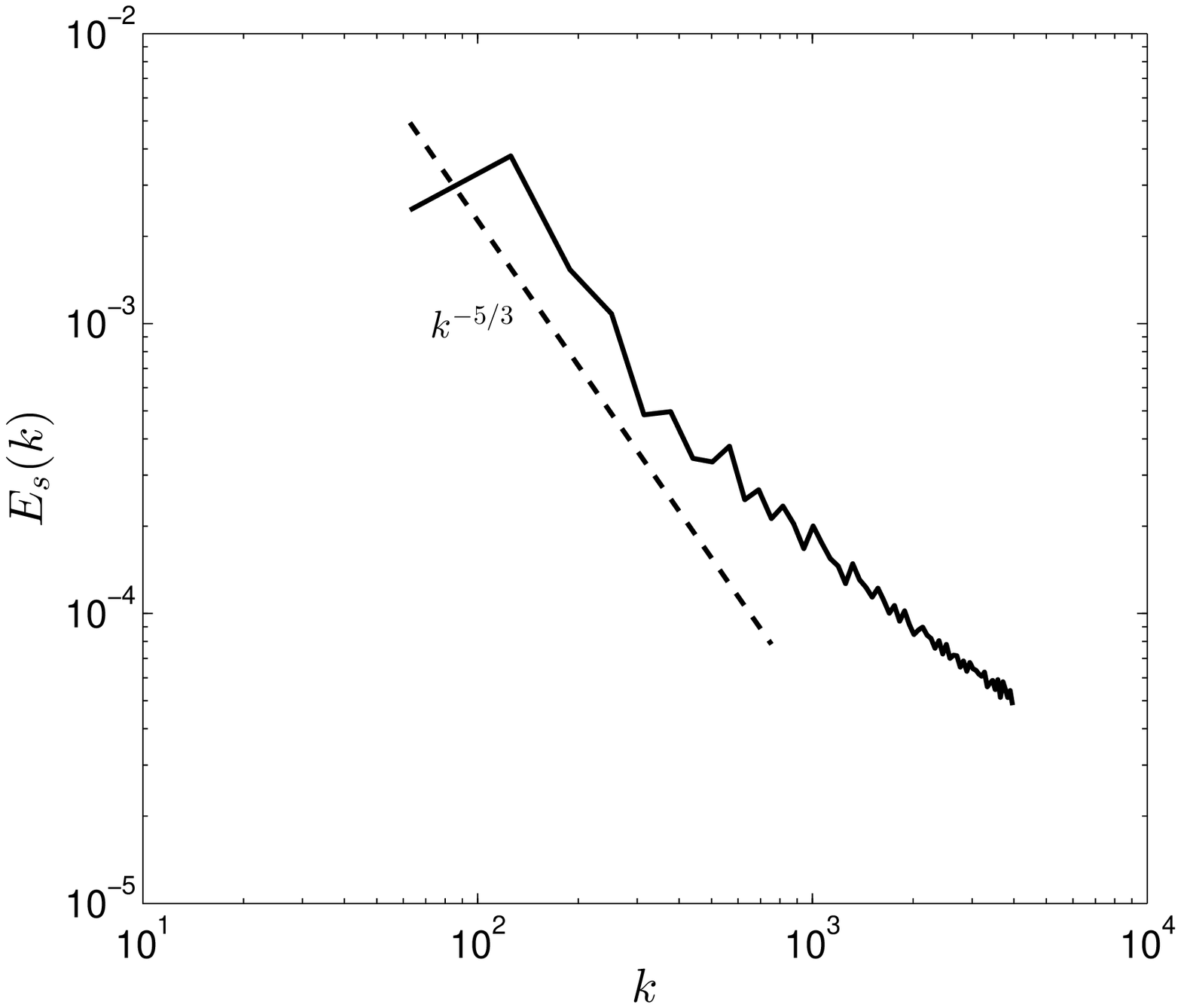} \\
\includegraphics[width=0.42\textwidth]{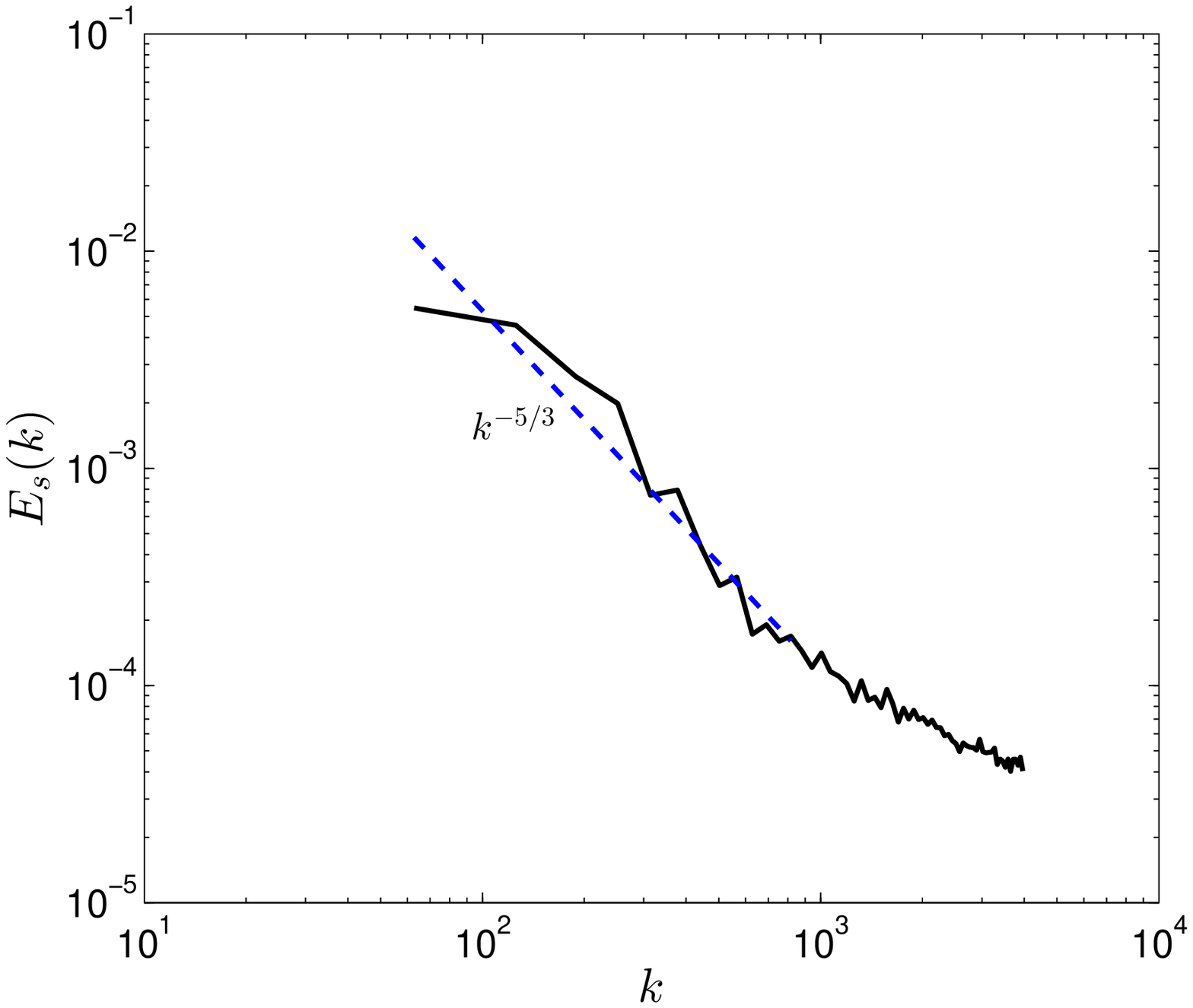}
\caption{
(color online). 
Energy spectrum $E(k)$ (arbitrary units) vs wavenumber $k$
($\rm cm^{-1}$) (time averaged over the saturated regime) 
as in Fig.~(\ref{fig:6}), but the simulations are
performed using the reconnection algorithm of Kondaurova et al.
\cite{Kondaurova}. Note that there is no significant difference from 
spectra obtained using our standard algorithm, see Fig.~(\ref{fig:6}).
Vortex tangles generated by uniform normal fluid (model 1, top),
synthetic normal fluid turbulence (model 2, middle) and
frozen Navier-Stokes turbulence (model 3, bottom).
The dashed lines indicate the $k^{-1}$ (top) and the $k^{-5/3}$
dependence (middle and bottom), respectively. Parameters: 
temperature $T=1.9~\rm K$, $V_n=0.75~\rm cm/s$ (for model 1), 
${\rm Re}= 81.59$ (for model 2), and ${\rm Re}= 3025$ (for model 3).}

\label{fig:7}
\end{center}
\end{figure}

Turbulent vortex tangles can be produced in the laboratory using
various means: by imposing a flux of heat, by pushing liquid helium~II
through pipes, or by stirring it with moving objects.
The numerical experiments presented here show that reporting
the vortex line density $L$ is not enough to characterize the
nature of the superfluid turbulence which can be generated in helium~II.
Vortex tangles with the same value of $L$ may have very
different energy spectra, depending on the normal fluid flow which
feeds energy into the vortex lines, as shown in Fig.~\ref{fig:6}.
If the normal fluid is turbulent, energy is contained in the large eddies
and is distributed over the length scales consistent with the classical 
$k^{-5/3}$ Kolmogorov law at large $k$, 
suggesting the presence of a Richardson cascade.
If the normal fluid is uniform, most of the energy is contained
at the intermediate length scales, and the energy spectrum scales 
consistently with $k^{-1}$ at large $k$. 
Using a terminology already in the literature, we identify these two forms
of superfluid turbulence as `Kolmogorov tangles' and `Vinen tangles'
respectively.

The superfluid velocity field is determined by the instantaneous
configuration of vortex lines. Since the superfluid velocity field
decays only as $1/r$ away from the axis of a quantum
vortex line, the interaction
between vortex lines is long-ranged, at least in principle.
By examining the ratio of local and nonlocal contributions to the
total velocity field, we have determined that in Vinen tangles far-field 
effects tend to cancel out ($v_s^{non}/v_s^{self} \approx 25 \%$
independently of $L$),  the motion of a vortex line is mainly
determined by its local curvature, and the vortex tangle is homogeneous. 
In Kolmogorov tangles, on the contrary,
nonlocal effects are dominant and increase with the vortex line density;
this stronger vortex-vortex interaction leads to the clustering of vortex 
lines, for which the vortex tangle is much less homogeneous and contains
coherent vorticity regions, in analogy to what happens in ordinary turbulence.
The presence of intermittency effects such as
coherent structures in the driving normal fluid,
which we have explored with model 3,  enhances the formation of superfluid
vortex bundles, resulting in a more inhomogeneous superfluid energy
and in larger nonlocal contributions to the vortex lines' dynamics.

Future work will explore the problem for turbulence
with nonzero mean flow.

\begin{acknowledgments}
We thank Kalin Kanov for help with DNS data.
\end{acknowledgments}

\newpage

%%%%%%%%%%%%%%%%%%%%%%%%%%%%%%%%%%%%%%%%%%%%%%%%%%%%%%%%%%
% FIG 1

%%%%%%%%%%%%%%%%%%%%%%%%%%%%%%%%%%%%%%%%%%%%%%%%%%%%%%%%%%%
% FIG 2

%%%%%%%%%%%%%%%%%%%%%%%%%%%%%%%%%%%%%%%%%%%%%%%%%%%%%%%%%%%
% FIG 3

%%%%%%%%%%%%%%%%%%%%%%%%%%%%%%%%%%%%%%%%%%%%%%%%%%%%%%%%%%%
% FIG 4

%%%%%%%%%%%%%%%%%%%%%%%%%%%%%%%%%%%%%%%%%%%%%%%%%%%%%%%%%%%
%% FIG 5

%%%%%%%%%%%%%%%%%%%%%%%%%%%%%%%%%%%%%%%%%%%%%%%%%%%%%%%%%%
% FIG 6

%%%%%%%%%%%%%%%%%%%%%%%%%%%%%%%%%%%%%%%%%%%%%%%%%%%%%%%%%%
% FIG 7

%%%%%%%%%%%%%%%%%%%%%%%%%%%%%%%%%%%%%%%%%%%%%%%%%%%%%%%%%%
% FIG 8

%%%%%%%%%%%%%%%%%%%%%%%%%%%%%%%%%%%%%%%%%%%%%%%%%%%%%%%%%%
%% FIG 9

%%%%%%%%%%%%%%%%%%%%%%%%%%%%%%%%%%%%%%%%%%%%%%%%%%%%%%%%%%%
% FIG 10

%%%%%%%%%%%%%%%%%%%%%%%%%%%%%%%%%%%%%%%%%%%%%%%%%%%%%%%%%%%
% FIG 11

%%%%%%%%%%%%%%%%%%%%%%%%%%%%%%%%%%%%%%%%%%%%%%%%%%%%%%%%%%
%%%%%%%%%%%%%%%%%%%%%%%%%%%%%%%%%%%%%%%%%%%%%%%%%%%%%%%%%%%%%
\end{document}